# Cosmological Constraints from Rotation Curves of Disk Galaxies

Julio F. Navarro[1]

Steward Observatory, University of Arizona, Tucson, AZ 85721.

**Abstract.** High-resolution N–body simulations show a remarkable similarity in the structure of dark matter halos formed through dissipationless hierarchical clustering. Independent of halo mass, power spectrum, and cosmological parameters, the density profiles of dark halos can be accurately fit by scaling a simple two-parameter profile. The two parameters of the fit, which can be expressed as the mass and characteristic density of each halo, are strongly correlated. The characteristic density is directly proportional to the mean density of the universe at the time of assembly of each halo. Low mass halos have higher characteristic densities, reflecting their higher collapse redshift. Using this halo structure to fit the rotation curves of a large sample of disk galaxies, we show that halos formed in the standard cold dark matter (CDM) scenario are too centrally concentrated to be consistent with the rotation curves of low-surface brightness galaxies. Less concentrated halos, such as those formed in a low-density, flat CDM universe ($\Omega_0 = 0.3$, $\Lambda = 0.7$), are favored by these observations. Our fits also indicate that the dark halo is important at all radii, and that its total mass is strongly correlated with the luminosity of the galaxy. This model provides a natural explanation for the disk-halo conspiracy and for the lack of dependence of the Tully-Fisher relation on galaxy surface brightness.

## 1. Introduction

Ever since Gunn & Gott (1972) showed that the structure of systems formed through gravitational collapse may contain clues to the cosmological parameters, the structure of dark matter halos formed in different cosmogonies has been the subject of numerous studies. The analytic work of Fillmore & Goldreich (1984), Bertschinger (1985) and Hoffman & Shaham (1985) suggested that the equilibrium mass profiles of dark halos should be well approximated by power laws, and that the power-law slope should depend sensitively on the cosmological parameters. These results were very influential in the interpretation of subsequent numerical studies, and prompted many authors to fit power-laws to the results of N–body simulations (Quinn et al 1986, Frenk et al 1988, Efstathiou et al 1988, Zurek, Quinn & Salmon 1988, Crone et al 1994). The general trends predicted by the analytic studies were generally confirmed, although significant deviations from power-laws were also reported. Such deviations were confirmed by the work of Dubinski & Carlberg (1991) and Navarro, Frenk & White (1995), who found that CDM halos were best described by a density profile with a gently changing logarithmic slope rather than a single power law.

---

[1]Bart J. Bok Fellow

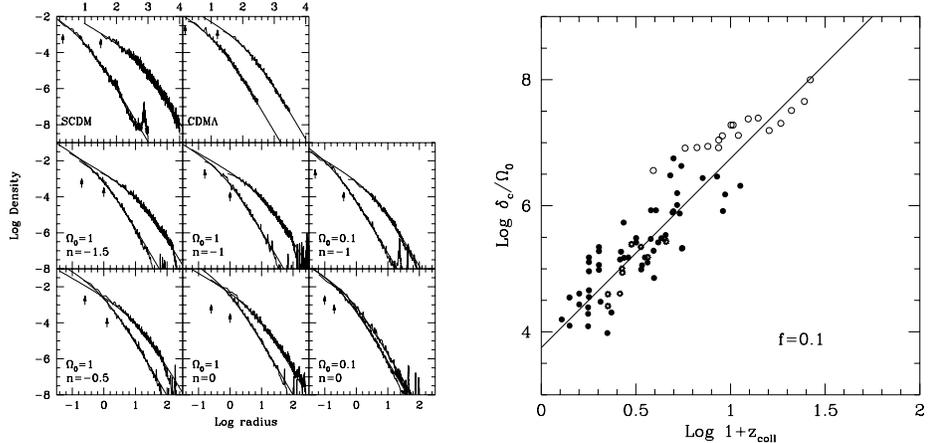

Figure 1. (a) Density profiles of dark matter halos. (b) Characteristic density vs collapse redshift for all simulated halos measured directly in the simulations. Filled circles refer to halos in $\Omega_0 = 1$ universes, open circles correspond to those in open models and starred symbols to the CDM$\Lambda$ model. The solid line shows the dependence predicted by eq. 2.

Follow-up work by Navarro, Frenk & White (1996a), Cole & Lacey (1996), and Tormen, Bouchet & White (1996) confirmed this result and showed that the simple two parameter profile,

$$\frac{\rho(r)}{\rho_{crit}} = \frac{\delta_c}{(r/r_s)(1 + r/r_s)^2}. \qquad (1)$$

provides an accurate fit to the structure of dark matter halos formed in an Einstein-de Sitter universe, regardless of halo mass and power spectrum. In eq. 1, $\rho_{crit} = 3H_0^2/8\pi G$ is the critical density for closure, $\delta_c$ is a (dimensionless) characteristic density, and $r_s$ is a scale radius. We write Hubble's constant as $H_0 = 100\,h\,\mathrm{km\,s^{-1}\,Mpc^{-1}}$ in this contribution.

We report here on the results of further N–body simulations that extend these results to low-density, flat and open cosmogonies. In total, we have analyzed the structure of halos in 8 different cosmological scenarios. Five correspond to different power spectra in an Einstein-de Sitter ($\Omega_0 = 1$) universe; the standard biased CDM spectrum (SCDM model: $\Omega_0 = 1$, $h = 0.5$, $\sigma_8 = 0.63$), plus four power-law ($P(k) \propto k^n$, "scalefree") spectra with indices $n = 0, -0.5, -1$, and $-1.5$. Two further models correspond to power-law spectra ($n = 0$ and $-1$) in an open universe ($\Omega_0 = 0.1$). The last model we consider is a low-density CDM model with a flat geometry (CDM$\Lambda$: $\Omega_0 = 0.25$, $\Lambda = 0.75$, $h = 0.75$, $\sigma_8 = 1.3$). (Here and throughout this paper we express the cosmological constant $\Lambda$ in units of $3H_0^2$, so that a low-density universe with a flat geometry has $\Omega_0 + \Lambda = 1$.) Further details on the simulations can be found in Navarro, Frenk & White (1996b).

## 2. Mass Profiles

Figure 1a shows, for each cosmological model and at $z = 0$, the spherically averaged density profiles of one of the least and one of the most massive halos in each series. These halos span four orders of magnitude in mass in the case of the CDM models and about two orders of magnitude in mass in the case of the power-law runs. Radial units are kpc in the CDM panels (scale at top), and are arbitrary in the power-law runs. Density is in units of $10^{10} M_\odot/\text{kpc}^3$ in the CDM panels and in arbitrary units in the rest. The density profiles have been fit to the configuration of the halo closest to dynamical equilibrium between $z = 0.05$ and $z = 0$. Solid lines are fits to each halo profile using eq. 1. Clearly, this simple formula provides a good fit to the structure of all halos over about two decades in radius, from the gravitational softening radius (indicated by arrows in Figure 1a) to about the virial radius of each halo. The quality of the fit is essentially independent of halo mass or cosmological model, and implies a remarkable similarity of structure between dark matter halos formed in different hierarchically clustering scenarios.

We define the mass of a halo, $M_{200}$, as that of a sphere with mean interior density equal to $200\rho_{crit}$. The radius of this sphere, $r_{200}$, is usually called the "virial radius" of the halo. Equivalently, we can use the circular velocity at $r_{200}$ in order to characterize the mass of a halo, $V_{200} = (r_{200}/h^{-1}\text{kpc})$ km/s. There is a single free parameter in eq. 1 to fit halos of a given mass. This free parameter can be expressed either as the characteristic overdensity $\delta_c$ or as the "concentration" of the halo defined by $c = r_{200}/r_s$. Our models show that $M_{200}$ and $\delta_c$ (or $c$) are strongly correlated. We find that the characteristic density is simply proportional to the mean density of the universe at the time of collapse,

$$\delta_c(M) \propto \Omega_0(1 + z_{coll}(M, f))^3. \tag{2}$$

The collapse redshift, $z_{coll}(M, f)$, is defined as the time at which half the mass of the halo was assembled into clumps more massive than some fraction $f$ of its final mass. Defined this way, $z_{coll}$ can be computed for any power spectrum using a simple equation based on the Press-Schechter theory (see, eg., Lacey & Cole 1993),

$$\text{erfc}\left(\frac{\delta_{crit}(z_{coll}) - \delta_{crit}^0}{\sqrt{2(\Delta_0^2(fM) - \Delta_0^2(M))}}\right) = \frac{1}{2}, \tag{3}$$

where $\Delta_0^2(M)$ is the linear variance of the power spectrum at $z = 0$ smoothed with a top-hat filter of mass $M$, $\delta_{crit}$ is the density threshold for spherical collapse, and $\delta_{crit}^0 = \delta_{crit}(z = 0)$ (see Navarro et al 1996b for details on the computation of all these quantities). Figure 1b shows that, for $f = 0.1$, this identification predicts accurately the dependence of $\delta_c$ on $z_{coll}$. Similar results are obtained for other values of $f$, as long as $f \ll 1$. (Using $f = 0.01$ in eq. 3, the proportionality constant in eq. 2 is $\sim 3 \times 10^3$ independent of the shape of the power spectrum.) Because low-mass systems collapse earlier in these hierarchical clustering scenarios, this implies that low mass systems are more centrally concentrated (or denser) than their more massive counterparts.

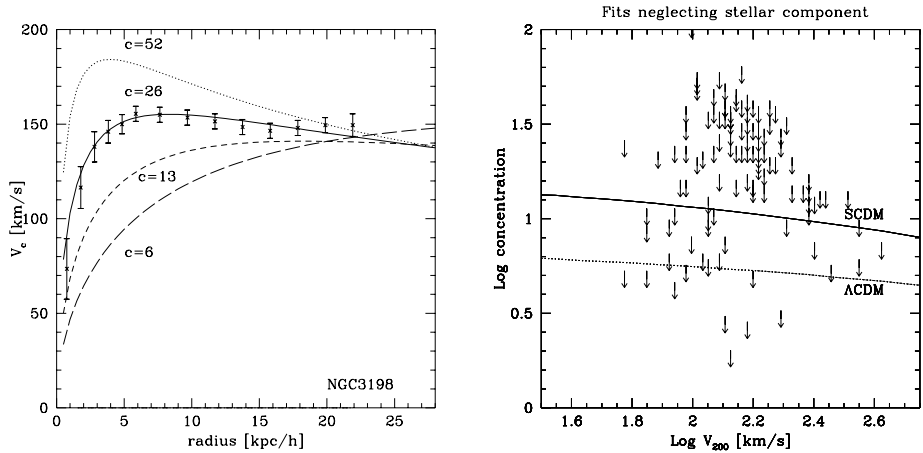

Figure 2. (a) Circular velocity profiles of halos with different concentrations and $V_{200} \sim 120$ km s$^{-1}$. Points with error bars correspond to NGC3198, plotted for reference. (b) Upper bounds on the concentration of the halos of a large sample of disk galaxies, derived by fitting $V_{200}$ and $c$ directly to the measured rotation curve of each galaxy, neglecting the contribution of the disk.

## 3. Rotation curves of disk galaxies

The circular velocity corresponding to the density profile of eq. 1 rises near the center as $r^{1/2}$ (where $\rho(r) \propto r^{-1}$), levels off and reaches a maximum at $r \sim 2\,r_s = 2\,r_{200}/c$ and declines beyond that radius. The concentration of a halo, therefore, regulates the shape of the circular velocity curve; high concentration corresponds to circular velocity profiles that rise fast near the center and low concentration systems are characterized by a slowly rising rotation curve. Figure 2a shows this explicitly in comparison with the rotation curve measured by Begeman (1987) for NGC3198. Note that concentrations of order $c \sim 26$ can produce a reasonable fit to the rotation curve of this galaxy (neglecting the disk). Higher concentrations predict too much mass near the center. Because we have neglected the contribution of the luminous disk and HI gas to the rotation curve, it is clear that $c \sim 26$ constitutes, for this galaxy, a firm *upper* limit to the concentration of the halo.

This procedure can be applied to any disk galaxy with a measured rotation curve in order to derive upper limits to the halo concentrations. We have collected from the literature a large sample of rotation curves spanning a large range in luminosity, rotation speed, and disk surface brightness. In particular, we have used the rotation curves of Matthewson et al (1992), as compiled by Persic & Salucci (1995, their sample $A$), Begeman (1987), Broeils (1992), de Blok et al (1996) and a few others. (For a complete list, see Navarro 1996.) This sample consists of about one hundred galaxies spanning more than two orders of magnitude in surface brightness, almost three orders of magnitude in total luminosity, and rotation speeds that range between 50 and 300 km/s.

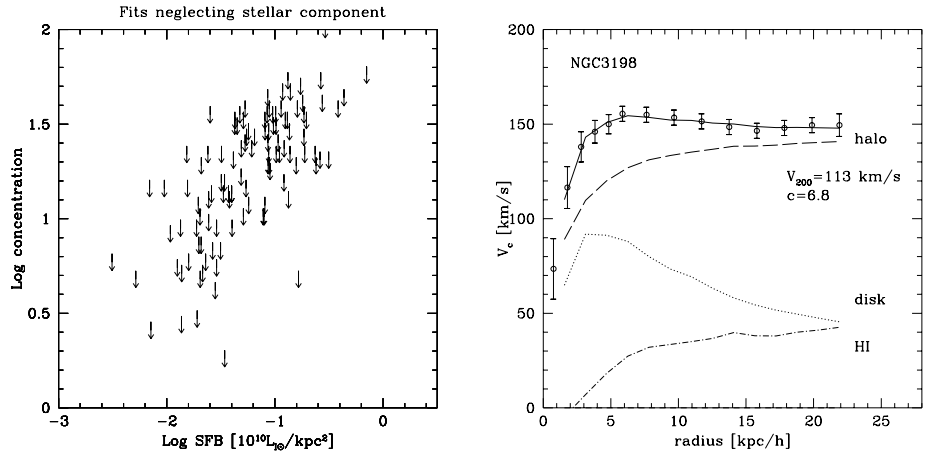

Figure 3. (a) Same upper bounds on $c$ shown in Fig. 2b, plotted as a function of the galaxy's surface brightness. (b) Halo-disk fit to the rotation curve of NGC3198.

Figure 2b shows the upper limits on halo concentration derived from fits such as those shown in Figure 2a. The solid line in Figure 2b indicates the loci corresponding to halos formed in the standard CDM scenario. A large number of galaxies lie below this line, indicating that SCDM halos are too centrally concentrated to be consistent with their rotation curves. The dotted line labeled "$\Lambda$CDM" corresponds to a low-density $\Omega_0 = 0.3$, $\Lambda = 0.7$, $h = 0.5$, CDM model normalized to match the amplitude of the microwave background fluctuations measured by COBE ($\sigma_8 \sim 0.7$, we note that this is only slightly lower than the $\sigma_8 \sim 0.8$ required to match the observed abundance of rich clusters, see Eke et al 1996). It is clear that the observations favor the lower concentration halos expected in the $\Lambda$CDM scenario.

The galaxies whose rotation curves rise too slowly compared with the SCDM model are predominantly low surface brightness galaxies, as shown in Figure 3a. Here we plot the surface brightness of a galaxy in the $I$-band, defined as $SFB = L_I/r_{disk}^2$, where $L_I$ is the total $I$-luminosity of the galaxy and $r_{disk}$ is the exponential scalelength of the disk. There is a clear correlation between the maximum concentration of the halo and galaxy surface brightness. This reflects the fact that low surface brightness (LSB) galaxies tend to have slowly rising rotation curves while high-surface brightness (HSB) galaxies have rotation curves that rise fast and stay flat or even decline slightly in the outer parts.

The data in Figure 3a show that the conclusions of Flores & Primack (1994) and Moore (1994), who pointed out that SCDM halos were too concentrated to be consistent with the rotation curves of a few dwarf irregulars such as DDO154, DDO160, or DDO105, actually apply to most low-surface brightness galaxies, regardless of their total luminosity or rotation speed. Indeed, our sample contains dwarf low surface brightness galaxies such as the DDO galaxies mentioned above, but also giant LSB galaxies such as UGC6614, whose peak rotation speed reaches $\sim 200$ km/s.

We emphasize that the correlation shown in Figure 3a does not imply that the halos surrounding HSB galaxies are more concentrated than those surrounding LSBs. Rather, it indicates the importance of the disk in shaping the circular velocity profile of the system. Halos with concentrations as low as those predicted by the $\Lambda$CDM model in Figure 2b ($c \approx 5$, almost independent of halo mass) have circular velocity profiles that would be slowly rising over the radial range corresponding to the typical luminous radii of galaxy disks (see Figure 2a). *Flat* rotation curves are therefore a direct indication of the disk's overall contribution to the gravitational potential of the galaxy. The rotation curves of HSBs are flat because the disk is important near the center while those of LSBs trace more closely the (slowly rising) circular velocity profile of the halo.

If this idea is correct the halos of all disk galaxies, irrespective of surface brightness, should have concentrations as low as those expected in the $\Lambda$CDM model. To test this idea, we attempted to fit the rotation curves in our sample allowing three parameters to vary: the halo circular velocity, $V_{200}$, the halo concentration, $c$, and the mass-to-light ratio of the stellar disk, $(M/L)_{disk}$. The fitting procedure includes, when possible, the contribution of neutral hydrogen (HI) and assumes that halos respond adiabatically to the presence of the disk (see Navarro et al 1996a for details on the adiabatic halo response).

This procedure works well only in galaxies with very extended rotation curves. This is hardly surprising, since it is clear that only rotation curves that sample a good fraction of the virial radius of the halo are capable of constraining well the halo parameters. A simple estimate of the virial radius assumes that the halo circular velocity is equal to the (maximum) rotation speed of the disk, $r^e_{200} = (V_{max}/\text{km s}^{-1})h^{-1}$ kpc. Empirically, we find that the procedure returns relatively tight constraints only when applied to sixteen galaxies whose rotation curves extend out to radii $\gtrsim 0.1 r^e_{200}$. These are all late type spirals with HI rotation curves that extend well beyond the optical disk and that span a wide range in surface brightness.

Figure 3b shows the result of this exercise in the case of NGC3198, the same galaxy illustrated in Figure 2a. Even in this case, with data extending out to more than $\sim 20$ disk scalelengths, the uncertainty in the recovered parameters is large. Figure 4 shows the fit parameters for the sixteen galaxies. The error bars represent 1-$\sigma$ uncertainties in each of the parameters. Despite the large uncertainty this exercise yields a few interesting results. (i) The concentration parameter of halos surrounding these galaxies is roughly independent of surface brightness or disk rotation speed. (ii) The maximum rotation speed of the disk ($V_{max}$) appears to be a good indicator of the circular velocity of the halo at the virial radius ($V_{200}$). Because of the Tully-Fisher relation (see Figure 6) this implies that the total luminosity of a galaxy is a good indicator of the total mass of the halo. (iii) The stellar mass-to-light ratios cluster at about $\sim 1h\,(M/L)_\odot$, but values as high as $\sim 6h$ are likely. There is little evidence of a strong correlation between disk $M/L$ ratios and surface brightness. These conclusions are similar to those of Persic & Salucci (1991) and Persic, Salucci & Stel (1996), although our analysis differs significantly from theirs.

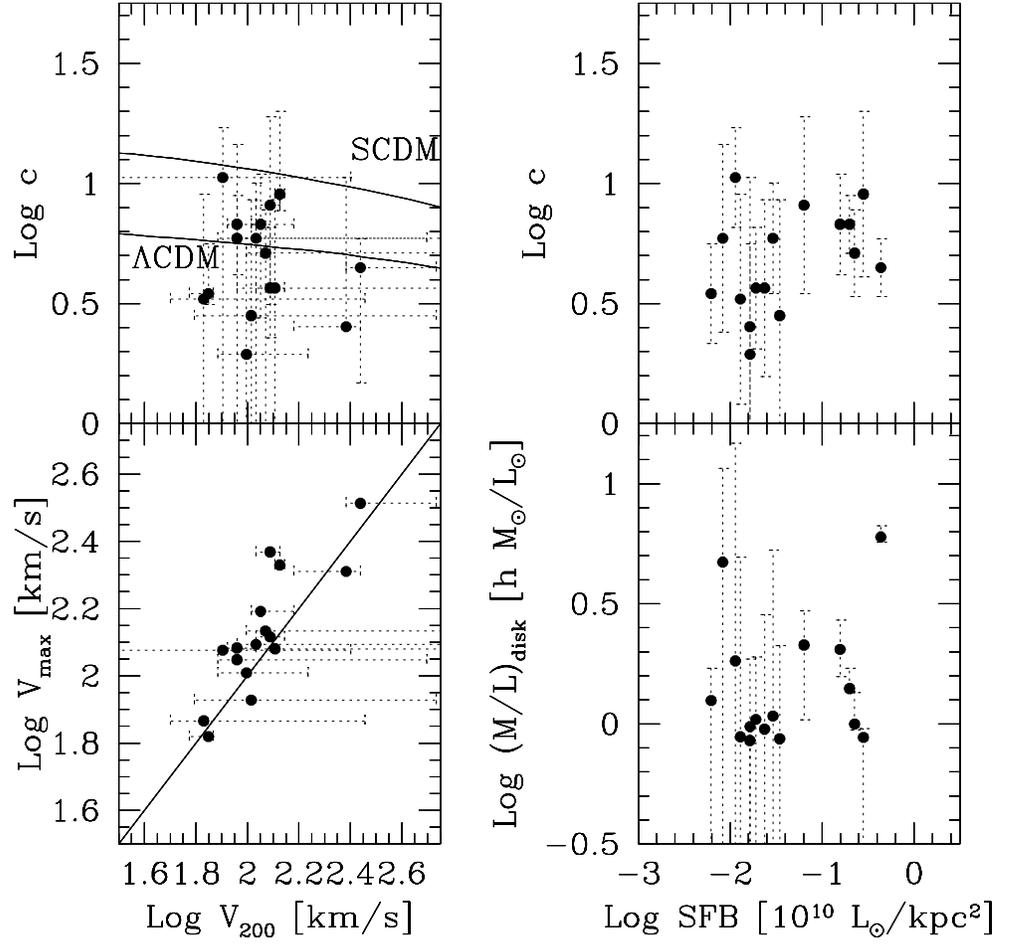

Figure 4. Results of three-parameter ($V_{200}$, $c$, and $(M/L)_{disk}$) fits to the rotation curves of 16 late-type galaxies with extended rotation curves.

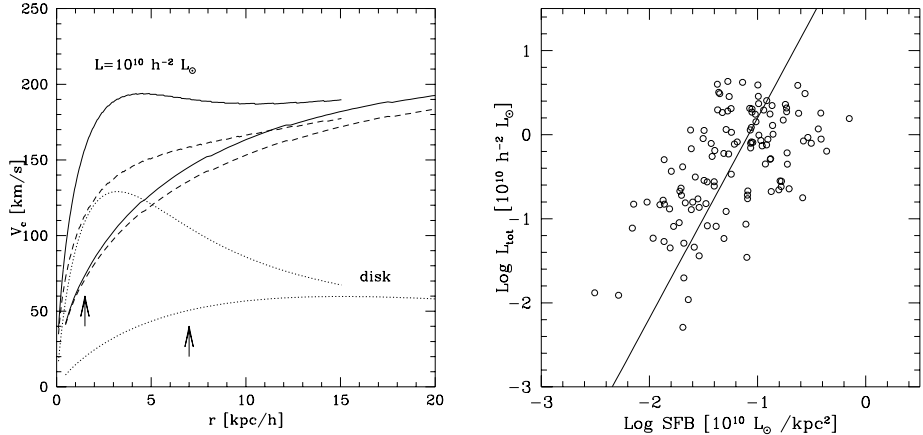

Figure 5. (a) Model rotation curves (solid lines) of two galaxies of the same luminosity but different surface brightness. Parameters are given in the text. Arrows indicate the exponential scalelength of each disk. Disk contributions are shown by dotted lines, halo contributions by dashed lines. Rotation velocities converge to the halo circular velocity beyond $\sim 3\,r_{disk}$. (b) Total I-band luminosity vs mean surface brightness for galaxies in our sample.

The solid lines in Figure 5a show the predictions of this model for the rotation curves of two galaxies of the same luminosity, $L_I = 10^{10} h^{-2} L_\odot$, but very different surface brightness. One galaxy in this case has $r_{disk} = 1.5\,h^{-1}$ kpc, and the other $r_{disk} = 7\,h^{-1}$ kpc (indicated by the arrows in the figure). This corresponds to a range of more than one order of magnitude in surface brightness, consistent with the full width of the surface brightness distribution (at this luminosity) for galaxies in our sample. Both galaxies are assumed to be surrounded by the same halo, with $V_{200} \sim 200$ km s$^{-1}$, and concentration $c = 5$. We have assumed in both cases that the disk has $(M/L)_{disk} = 1.5\,h\,(M/L)_\odot$, and that the halo responds adiabatically to the presence of the disk. The shapes of the rotation curves of these two galaxies are significantly different, consistent with the trend shown in Figure 3a.

Figure 5a also illustrates a couple of interesting points. In the case of an HSB the circular speed peaks at about $\sim 2\,r_{disk}$, coincident with the peak in the disk contribution to the rotation curve. This is one aspect of the "disk-halo conspiracy", ie. the puzzling coincidence between the shapes of the observed inner rotation curve and that expected from the disk. This result has been interpreted as favoring the "maximal disk" hypothesis (see Broeils & Courteau, this volume) but arises naturally in our models although the disk contribution is far from "maximal". Indeed, the halo contributes about half of the circular velocity within this radius and dominates further out, where the rotation curve remains flat. Note also that, because of the adiabatic halo response, the transition between the region where the disk is important to that where the halo dominates is smooth and featureless. The situation is different in the case of LSB galaxies, where the halo dominates at all radii.

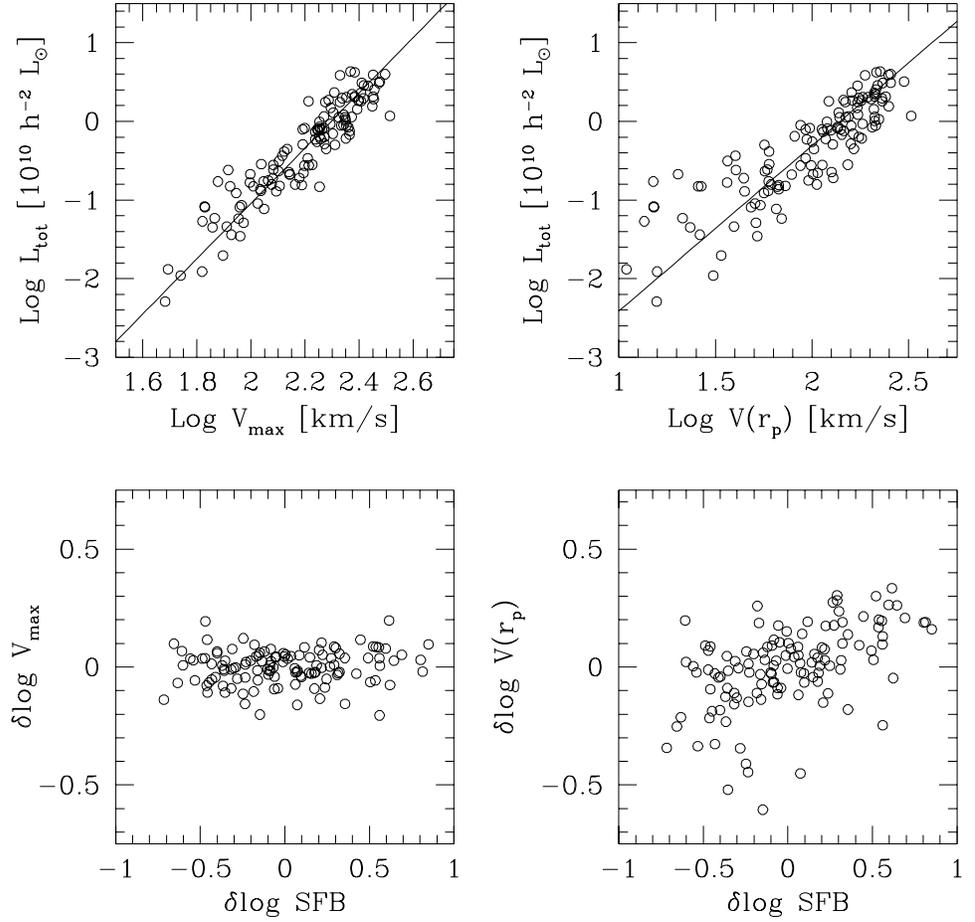

Figure 6. Upper panels: Total I-band luminosity vs rotation speed. $V_{max}$ is the maximum rotation speed measured in the rotation curve, and $V(r_p)$ is the rotation speed at a physical radius independent of surface brightness, $r_p = 3\,(V_{max}/200 \text{ km s}^{-1})\,h^{-1}$ kpc. Solid lines are least-squares fits. Lower panels: Velocity residuals (from upper panels) vs surface brightness residuals, at a given luminosity. The latter are computed from power-law fits to the $L_{tot}$-SFB relation (Figure 5b).

The rotation velocity of the two galaxies in Figure 5a is similar only at $r \gtrsim 3\,r_{disk}$. Measured at a fixed physical radius, e.g at $3\,h^{-1}$ kpc, the rotation speeds of these two galaxies would differ by more than 60%. One consequence of our modeling is, therefore, that the Tully-Fisher relation of LSB's and HSB's would be significantly different if velocities were measured at a radius comparable to the disk scalelength of an HSB, but should be equivalent when measured at radii larger than a few *individual* scalelengths. This effect is clearly seen in our sample. Measuring rotation speeds at a radius of $r_p = 3\,(V_{max}/200$ km s$^{-1})h^{-1}$ kpc results in a significant offset between the zero-point of the Tully-Fisher relation of HSBs and LSBs. This is illustrated in Figure 6, where we show the "Tully-Fisher" relations for our sample, constructed using either $V_{max}$ or $V(r_p)$. The lower-right panel shows that at a given luminosity, higher than average surface brightness galaxies tend to have higher than average rotation speeds. This offset disappears once maximum rotation speeds within $\gtrsim 3\,r_{disk}$ (the typical extent of the rotation curves in our sample) are used to construct the Tully-Fisher relation (left panels in Figure 6). The weak dependence of this relation on surface brightness (Zwaan et al 1995) is, therefore, a direct result of the fact that velocity measurements typically extend beyond $\sim 3\,r_{disk}$.

## 4. Summary

The mass profiles of dark matter halos formed through dissipationless hierarchical clustering seem to have a universal shape which is independent of halo mass, power spectrum of initial density fluctuations, and cosmological parameters. Halos of a given mass are completely specified by their characteristic overdensity, which simply reflects the mean density of the universe at the time of their collapse. Since collapse times change when the power spectrum or the cosmological parameters are varied, halo mass profiles can be used to derive strong constraints on cosmogonical models.

The shape of the rotation curves of disk galaxies is consistent with this mass profile. However, halos as dense as expected in the standard CDM model are strongly ruled out by the rotation curves of low surface brightness disk galaxies. Less concentrated halos, such as those expected in a COBE-normalized low-density, flat CDM model, are favored by these observations.

Rotation curve fits using halos formed in this cosmogony indicate that the luminosity of a galaxy is, to first order, a good indicator of its surrounding halo's circular velocity (or mass). This model reproduces the systematic difference in rotation curve shape as a function of surface brightness. HSB's have fast-rising, flat rotation curves because of the important contribution of the disk to the overall potential. On the other hand, the rotation curves of LSB's rise slowly and are, on average, fairer tracers of the dark matter potential. This is a "natural" prediction of our modeling, and does not assume any radical dependence between stellar $M/L$ ratios and surface brightness, or any systematic difference between the structure of halos surrounding LSBs and HSBs.

The dark halo is important at all radii in all galaxies; in high surface brightness galaxies it typically contributes about half the total circular velocity at about a couple of disk scalelengths. This model also predicts that the Tully-Fisher relation should be independent of surface brightness *provided* that veloc-ity measurements extend at least to about 3 exponential disk scalelengths in all cases. If rotation speeds are measured at a fixed physical radius, e.g. at about the exponential scalelength of an HSB, the residuals of the Tully-Fisher relation correlate strongly with surface brightness, an effect that disappears if maximum velocities within $\sim 3\, r_{disk}$ are used.

We conclude from the success of this model in explaining this various observations that the structure of halos surrounding disk galaxies is very similar to that of dark matter halos formed in a low-density CDM cosmogony.

**Acknowledgments.** I am grateful to Carlos Frenk and Simon White, for allowing me to present results of our joint collaboration, as well as to H.-W. Rix for providing insightful criticism. Special thanks to W.J.G. de Blok, P. Salucci, and T. Pickering for making available some of the data used here in electronic form.